%Paper: hep-th/9204058
%From: <PJESENAT%KRYSUCC1.BITNET@pucc.princeton.edu>
%Date: MON, 20 APR 1992 02:02 EXP
%Date (revised): MON, 27 APR 1992 04:33 EXP
%Date (revised): MON, 19 OCT 1992 06:57 EXP

%%%%%%%%%%%%%%%%%%%%%%%%%%%%%%%%%%%%%%%%%%%%%%%%%%%%%%%%%%%%%%%%%%%%%%
%%                                                                  %%
%%    The Algebraic Structures of Topological Yang-Mills Theory     %%
%%                                                                  %%
%%           Final Version: Errors are fixed and                    %%
%%               some parts are rewritten                           %%
%%                                                                  %%
%%                        Jae-Suk Park                              %%
%%                                                                  %%
%%%%%%%%%%%%%%%%%%%%%%%%%%%%%%%%%%%%%%%%%%%%%%%%%%%%%%%%%%%%%%%%%%%%%%
%-----------------------------------------------
% This is a TeX files.
% You need Harvmac.
%-----------------------------------------------
\input harvmac
%
%-------------------Macros----------------------------------------
%

\def\be{\beta}
\def\ga{\gamma}
\def\de{\delta}
\def\la{\lambda}

\def\z{\omega}
\def\rd{\partial}
\def\dt{\de_{\!{}_{T}}}
\def\dw{\de_{\!{}_{W}}}
\def\db{\de_{\!{}_{B\!R\!S}}}

\def\z{\omega}
\def\p{\partial}
\def\CA{{\cal A}}

\def\CC{{\bf{\cal C}}}
\def\CF{{\cal F}}
\def\CW{{\cal W}}
\def\CU{{\cal U}}
\def\CM{{\cal M}}
\def\CG{{\cal G}}
\def\CUG{{\cal U/G}}
\def\hF{\hat{F}}
\def\hA{\hat{A}}
\def\hd{\hat{d}}
\def\hCF{\hat{\CF}}
\def\hCA{\hat{\CA}}
\def\bCC{{\bf\CC}}
\def\tW{\tilde{W}}
\def\tCW{\tilde{\CW}}

\def\tr{\hbox{Tr}\,}
\def\Fr#1#2{{#1\over#2}}
\def\f#1#2{{\textstyle{#1\over#2}}}

\def\Fr#1#2{{#1\over#2}}
\def\tr{\hbox{Tr}\,}

\def\rF#1{\Fr{\rd}{\rd #1}}

\def\ack{\bigbreak\bigskip\bigskip
\centerline{{\bf Acknowledgements}}\nobreak}
%
%----------------------References--------------------------------------------
%
\lref\W{
W.\ Witten, Comm.\ Math.\ Phys.\ 117 (1988) 353}
\lref\WW{
E.\ Witten, Introduction to cohomological field theories,
{\it in\/} Proc.\ Trieste Conference on
Topological methods in quantum field theories (ICTP, Trieste, June
1990), ed.\ W.\ Nahm et.\ al., (World Scientific, Singapore, 1991)
}
\lref\D{
S.K.\ Donaldson, J.\ Diff.\ Geo.\ 18 (1983) 269; 26 (1987) 397;
Topology 29 (1990) 257}
\lref\F{
A.\ Floer, Comm.\ Math.\ Phys.\ 118 (1988) 215}
\lref\A{
M.F.\ Atiyah, {\it in\/}\ The mathematical heritage of Herman Weyl,
eds.\ R.\ O.\ Wells Jr.\ et.\ al.,
Proc.\ Symp.\  Pure.\ Math.\ 48 (AMS, Providence, 1988)}
\lref\LP{
J.M.F. Labastida, M.\ Pernici, Phys.\ Lett.\ B212 (1988) 56}
\lref\BS{
L.\ Baulieu, I.M.\ Singer, Nucl.\ Phys.\ (Proc.\ Suppl.) B5 (1988) 12}
\lref\BMS{
R.\ Brooks, D.\ Montano, J.\ Sonnenschein, Phys.\ Lett.\ B214
(1988) 91}
\lref\H{
J.H.\ Horne, Nucl.\ Phys.\ B318 (1989) 22}
\lref\M{
R.\ Myers, Int.\ J.\ Mod.\ Phys.\ A5 (1990) 1369}
\lref\BRT{
D.\ Birminham, M.\ Rakowski, G.\ Thomson, Nucl.\ Phys.\ B315
(1989) 577}
\lref\BBRT{D.\ Birminham, M.\ Blau, M.\ Rakowski and G.\ Thomson,
Phys.\ Rep.\ 209 (1991) 129}
\lref\OT{
S.\ Ouvry, G.\ Thompson, Nucl.\ Phys.\ B344 (1990) 371}
\lref\K{
H.\ Kanno, Z.\ Phys.\ C43 (1989) 477}
\lref\OSB{
S.\ Ouvry, R.\ Stora and P.\ van Ball, Phys.\ Lett.\ B214 (1988) 223}
\lref\AS{
M.F.\ Atiyah and I.M.\ Singer, Proc.\ Nat.\ Acad.\ Sci.\ USA, 81
(1984) 2597}
\lref\BRS{
C.\ Becchi, A.\ Rouet and R.\ Stora, Comm.\ Math.\ Phys.\ 42
(1975) 127\semi I.V.\ Tyutin, Lebedev Inst.\ preprint, (1975)
\semi G.\ Curci and R.\ Ferrari.\ Nuovo Cimento, A32 (1976) 151;
A35 (1976) 273}
\lref\Stora{
R.\ Stora, Algebraic structure and topological origin of
anomalies, {\it in\/} Progress in gauge field theory, ed.\ G.\
't Hooft et al.\ (Plenum, N.Y., 1984)}
\lref\ZuminoA{
B.\ Zumino, Chiral anomalies and differential geometry, {\it
in\/} Relativity, group and topology II, eds.\ B.S.\ deWitt et.\
al., Les Houches 1983, (North-Holland, Amsterdam, 1984)}
\lref\ZuminoB{
B.\ Zumino, Nucl.\ Phys.\ B253 (1985) 477}
\lref\ZuminoC{
B.\ Zumino, Anomalies, cocycles and Schwinger terms,
{\it in\/} Symposium on anomalies, geometry and topology,
March 28-30, 1985,
eds.\ W.A.\ Bardeen and A.R.\ White (World Scientific, Singapore)}
\lref\MZ{
J.\ Manes, and B.\ Zumino, Non-triviality of gauge anomalies,
{\it in\/} Supersymmetry and its applications,
eds.\ G.W.\ Gibbons et.\ al., Proceedings of Nuffield Workshop
1985 (Cambridge Univ.\ Press, 1985)}
\lref\MSZ{
J.\ Manes, R.\ Stora and B.\ Zumino, Comm.\ Math.\ Phys.\ 102
(1985) 157}
\lref\BZ{
W.A.\ Bardeen and B.\ Zumino, Nucl.\ Phys.\ B244 (1984) 421}
\lref\ZWZ{
B.\ Zumino, Y.-S.\ Wu and A.\ Zee, Nucl.\ Phys.\ B239 (1984) 477}
\lref\AG{
L.\ Alverez-Gaum\'{e} and P.\ Ginsparg, Nucl.\ Phys.\ B243
(1984) 449 \semi L.\ Alverez-Gaum\'{e} and P.\ Ginsparg, Ann.\
Phys.\ 161 (1985) 423 }
\lref\GraAno{
L.\ Alvarez-Gaum\'{e} and E.\ Witten, Nucl.\ Phys.\ B 234 (1983)
269 \semi O.\ Alvarez, I.M.\ Singer and B.\ Zumino, Comm.\
Math.\ Phys.\ 96 (1984) 417 }
\lref\WZ{ J.\ Wess and B.\ Zumino, Phys.\ Lett.\ B37 (1971) 95}
\lref\MN{J.M.\ Maillet and A.J.\ Niemi, Phys.\ Lett.\ B223 (1989) 195}
\lref\Wu{S.\ Wu, Phys.\ Lett.\ B264 (1991) 339}
\lref\Gri{V.\ Gribov, Nucl.\ Phys.\ B139 (1978) 1}
\lref\S{I.M.\ Singer, Comm.\ Math.\ Phys.\ 60 (1978) 7}
%
%----------------------Title----------------------------------------
%
\Title{\vbox{\baselineskip12pt\hbox{ESENAT-92/04}\hbox{hep-th/9204058}}}
{Algebraic Structures}
\vskip -0.3in\centerline{{\titlefont
of Topological Yang-Mills Theory}}
\vskip 0.3in
\centerline{Jae-Suk Park\footnote{$^\dagger$}
{Bitnet: pjesenat@krysucc1}}

\bigskip
\centerline{{\it ESENAT}}
\centerline{{\it Research Institute for Theoretical Physics}}
\centerline{{\it 70-30 Changcheon}}
\centerline{{\it Seoul 120-180, Korea}}

\vskip 0.5in

\centerline{{\bf Abstract}}
\vskip 0.2in
We discuss the algebraic structure of the various BRST symmetries
associated with topological Yang-Mills theory as a generalization of the
BRS analysis developed for the non-Abelian anomaly in the local
Yang-Mills theory.
We show that our BRST algebra leads to an extended
{\it Russian formula\/} and {\it descent equations}, which contains
the descent equation of Yang-Mills theory as sub-relations.
We propose the non-Abelian anomaly counterpart in Topological
Yang-Mills theory using the extended descent equation.
We also discuss the geometrical structure of our BRST symmetry
and some explicit solutions of the extended descent equation
are calculated.

%\draft
\Date{April, 1992; Final Version: September, 1992}
%
%----------------------Body--------------------------------------
%\submit

\newsec{Introduction}

The topological Yang-Mills theory\W\ (TYM in short) is a field theoretical
interpretation of Donaldson's polynomial invariants of smooth four
manifolds\D\ and the first example of the topological field theories
(TFT's) in the cohomological nature.

Originally, Witten has constructed TYM  using a
relativistic generalization of Floer's theory of three manifolds\F\
following Atiyah's conjecture\A. He also proposed that the theory might be
obtained as a BRST quantized version of underlying  general covariant
theory with higher symmetry. It has been discovered that the  theory can be
obtained by a BRST quantization of the underlying action which is the Chern
class or simply zero\LP\BS\BMS.  The classical action has local gauge
symmetry as well as topological symmetry, and Witten's action can be
recovered if one quantizes the underlying action such that the
configuration space is the instanton moduli space, while preserving the
local gauge symmetry.

The remaining gauge symmetry should also be properly fixed, and there are
two different approaches. One is to introduce another BRST operator\foot{We
shall use the same terminology as ref.\K\ for $\db$, $\dw$ and $\dt$.}
which is equivalent to the conventional operator $\db$ in addition to
$\dw$\H. And the other is to fix the entire symmetry using a single BRST
operator $\dt$\LP\BS, which can be roughly decomposed as $\dt \sim \dw +
\db$. Although both approaches have some problems, it has been demonstrated
that they are equivalent\M. The algebraic and geometrical structure of
$\dt$ algebra have been analyzed in ref.\K\OSB.  Kanno\K\ in particular
has suggested that the correct geometrical
framework of $\dt$ algebra is the universal bundle\AS.

In the second approach, one identifies  the Faddev-Popov ghost with the ghost
$c$ of $\dt$ algebra, and regard $\dt$ as an unification of $\dw$ and $\db$.
In this paper, we suggest that this interpretation is somewhat
misleading and an alternative way is possible, which naturally
leads to an unified formalism of $\dw$ and $\db$ as well as $\dt$
algebras. The main step of our approach is to identify the correct BRS
sector among the generalized BRST symmetry relevant in TYM following
the traditional method dealing with the local Yang-Mills
theory\ZuminoA\ZuminoB\MSZ. Then, we suggest an extended
{\it Russian formula\/} and the corresponding extended
{\it descent equation}.

The Witten's observables for the Donaldson
polynomial invariant can be interpreted  as the Abelian anomaly
counterparts in TYM\MN\Wu. On the other hand,
it is well known that the Russian formula with the descent equation
of  the Yang-Mills theory is
an systematic algebraic method to find the $2n$-dimensional
non-Abelian anomaly from the Abelian anomaly of two high dimension.
Thus, our formalism can be served as an algebraic
method for studying the non-Abelian anomaly counterpart in TYM.
Historically, the first application of the universal bundle formalism
in physics was the geometrical interpretation of the non-Abelian
anomaly\AS\Stora.  It will be very interesting to investigate the
universal bundle formalism whether the structure of non-Abelian anomaly
survives in TYM.

In sect.\  2, we give a brief review of the algebraic structures of  $\db$
algebra\BRS\ and the anomalies of Yang-Mills theory, which are the starting
points of this article.  In sect.\  3, we discuss some difficulties of
$\dt$ algebra and propose a unified $\de$ algebra which can be decomposed
as $\de =\dt + \db$ following the method introduced in sect.\ 2. We  show
that $\db$ algebra is identical to the additional BRS algebra in Horne's
approach\H\M. Then, we give a extension of
the {\it Russian formula} which leads to an {\it extended descent equation}.
We shall see that the descent equations of Yang-Mills theory are
sub-relations of the extended descent equations,
which shows that TYM has surprisingly rich structures.
In sect.\ 4, we suggest
that our descent equation can be used to study the non-Abelian anomaly
counterpart in TYM.  We show that there is a rich class of $\db$
invariant quantities, which contains the consistent anomaly counterparts
in TYM.

The conclusion with some open questions is given in the
final section.
In appendix A we discuss the geometrical
origins of the various BRST symmetries and the extended Russian formula
based on the universal bundle formalism. This appendix
is largely complementary to sect.\ 3.
Some solutions of the extended descent equation are presented
in Appendix B.

\newsec{The BRS Algebra and Anomalies}

In this section, we will briefly review the algebraic structures of the BRS
algebra and the anomalies\foot{The materials in this section are all
standard and I have generally followed Zumino\ZuminoA\ZuminoB\ with the
same conventions.} for later use.  Let $M$
be a compact oriented $2n$-dimensional Riemann manifold and $P$ a principal
G-bundle over $M$.  We denote $\CU$ to the space of all connections on $P$
and $\CG$ to the gauge group, which is the bundle automorphism leaving a
base point in $P$.  Let $A$ denote a Lie algebra valued (Ad($P$)-valued)
vector potential (connection one-form) over $M$.
The curvature two-form $F$ is given by
\eqn\efs{F = dA + A^2,}
which satisfies the Bianchi identity
\eqn\eb{dF + [A,F] = 0.}
%Thus we can see that $\tr F^{n}$ is gauge invariant.
%From the Bianchi identity, we can get
%\eqn\edfz{d \tr F^n = 0.}
%By the Poincar\'{e} lemma, we can write
%\eqn\ecs{\tr F^{n} = d\z_{2n-1}{}^{0}(A,F)}
%where $\z_{2n-1}{}^0(A,F)$ is the Chern-Simons term.
%Under the gauge transformation the Chern-Simons term transforms
%as
%\eqn\ecsgt{
%\z_{2n-1}{}^0(A^{g},F^{g}) =
%\z_{2n-1}{}^0(A,F) + d\al_{2n-2} +
%\z_{2n-1}{}^0(g^{-1}dg,0),}
%where the last term is known as the Wess-Zumino term.
Under the gauge transformation
\eqn\egt{\eqalign{
&A \rightarrow  g^{-1}Ag + g^{-1}dg, \cr
&F \rightarrow  g^{-1}Fg,\cr}
}
where $g(x,\la)\in \CG$ is an element of the gauge group, which is a
function of the space-time variables $x^\mu$ and of additional parameters
$\la^i$ which specify the particular element of the gauge group.
Thus for a given
vector potential $A(x)$ transformed one $\CA$ defined by
\eqn\ecA{\CA \equiv g^{-1}Ag + g^{-1}dg,}
depends on the parameter $\la^{i}$. The gauge transformed curvature
$\CF$ is given by
\eqn\gtc{
\CF\equiv d\CA + \CA^2 = g^{-1}F\, g.
}

Now one can distinguish the exterior derivative in the direction of $x$
from that in the direction of the group such that
\eqn\edde{\eqalign{
&d= dx^{\mu}{\p \over \p x^{\mu} },\cr
&\db = d\la^i {\p \over \p\la^i}, \cr}
}
which satisfy
\eqn\enp{d^2=\db^2 = d\db +\db d = 0.
}
Though there was no given vector potential in the group direction,
it can be generated as the pure gauge, which is a Lie algebra valued
one-form (Maurer-Cartan form) in $\CG$
\eqn\ev{
v = g^{-1}\db g.
}
Then, it follows that
\eqn\ebrs{\eqalign{
\db \CA &= - dv - \{\CA,v\}\equiv -d_\CA v,\cr
\db v &= - v^2,\cr}
}
which is nothing but the BRS transformations with $v$ being the
Faddev-Popov ghost.

Then we can find that the gauge transformed field
strength $\CF$ satisfies the {\it Russian formula};
\eqn\erusA{
\CF \equiv d\CA +\CA^2
= (d + \db)(\CA + v) + (\CA + v)^2,
}
and the Bianchi identity\eqn\ebi{
d \CF + [\CA ,\CF] = (d +\db)\CF + [\CA + v, \CF] = 0.
}
Now we can define a symmetric invariant polynomial\foot{See appendix B
for convention.} of degree $n$
\eqn\trivial{
P(\CF^n)=P(F^n),
}
satisfying
\eqn\erusB{d\, P(\CF^n) = (d + \db ) P(\CF^n) = 0,}
which follows from \ebi. By the Poincar\'{e} lemma
\eqn\epl{
(d + \db)\z_{2n-1}(\CA + v, \CF)
= d \z_{2n-1}{}^{0}(\CA , \CF).
}
Expanding $\z_{2n-1}(\CA + v, \CF)$ to the power of $v$
\eqn\epv{
\z_{2n-1}(\CA + v, \CF) = \z_{2n-1}{}^{0}(\CA,\CF)
+\z_{2n-2}{}^{1} + \cdots + \z_{0}{}^{2n-1},
}
where the superscript indicates the power of $v$.
Using eq.\ \epl, we can get the descent equations
\eqn\edec{\eqalign{
d\z_{2n-2}{}^{1} + \db\z_{2n-1}{}^{0} &= 0, \cr
d\z_{2n-3}{}^{2} + \db\z_{2n-2}{}^{1} &= 0, \cr
\vdots &            \cr
d\z_{0}{}^{2n-1} + \db\z_{1}{}^{2n-2} &= 0, \cr
\db\z_{0}{}^{2n-1} &= 0. \cr  }
}
The second relation of the above set of equation  is
the Wess-Zumino consistency condition\WZ
\eqn\ewz{d\z_{2n-3}{}^{2} + \db\z_{2n-2}{}^{1} = 0,
}
that is,
\eqn\enaan{\int_{M} \z_{2n-2}{}^{1},
}
gives the non-Abelian anomaly in the $2n-2$ dimension ($\hbox{dim}(M)=
2n-2$).

The descent equation \edec\ is a systematic algebraic method to get
non-Abelian anomalies in $2n-2$ dimension from a $2n$ dimensional Abelian
anomaly, which is given by the Atiyah-Singer index theorem;
\eqn\eaan{
n_{+}-n_{-}=\int{1\over n!}\left({i\over 2\pi}\right)^n \tr \CF^n,
}
where $n_{+}$ ($n_{-}$) is the number of fermion zero modes of positive
(negative) chirality. The non-Abelian anomaly is normalized with an
additional factor of $2\pi$\ZuminoA\ such that
\eqn\ennano{{1\over n!}{i^n\over(2\pi)^{n-1}}\z_{2n-2}{}^1,
}
gives the non-Abelian anomaly in the $2n-2$ dimension.  In ref.\AG,
the non-Abelian anomaly has been derived from the families of index
theorems of two higher dimensions. They have also discussed the topological
origin of non-Abelian anomalies.  The general solutions of the descent
equation \edec\ can be obtained by various methods\ZuminoA\ZuminoB\MSZ\ZWZ.

\newsec{BRST Algebras of Topological Yang-Mills theory}
\subsec{Motivations}
Kanno has been suggested\K\ that that the natural geometrical framework of
$\dt$ algebra of \LP\BS\ is the universal bundle
constructed by Atiyah-Singer\AS. It is possible to calculate the total
curvature of the universal bundle over $M\times\CU/\CG$ and the
corresponding invariant polynomials, one of which leads to Witten's
observables in TYM theory for Donaldson's invariants. However, $\dt$
algebra is not identical to the universal bundle formalism up to the terms
which can be regarded as being similar to the usual BRS transformation.
It has been suggested\BRT\BBRT\ that $\dt$ algebra does not come from the
universal bundle but from a fullback bundle over $M\times \CU$.

Following \BBRT, consider a principal $G$-bundle $Q$ over $M\times \CU$.
Then one can decompose an arbitrary fixed connection (Lie algebra ${\bf
g}$-valued) one-form $\hA$ into
\eqn\ehA{
\hA = A + c,
}
where $A$ denotes an $(1,0)$-form tangent to $M$ and $c$ denotes
$(0,1)$-form tangent to $\CU$. Similarly, we can decompose arbitrary Lie
algebra valued $r$-form into $(p,q)$-form
\eqn\edcom{
A^{r}({\bf g}) = \sum_{r = p+q} A^{p,q}({\bf g}),
}
where $ A^{r}({\bf g})$ and $A^{p,q}({\bf g})$ denote the space
of ${\bf g}$-valued $r$-forms and $(p,q)$-forms, respectively.
We can also decompose the exterior derivative $\hd$ into
\eqn\ehd{
\hd = d + \de, \qquad
d:A^{p,q} \rightarrow A^{p+1,q}, \qquad
\de:A^{p,q}\rightarrow A^{p,q+1}.
}
{}From $\hd^2 = 0$, we get
\eqn\enpo{
d^2 = \de^2 = d\de + \de d = 0.
}
The curvature two-form $\hF$ is decomposed as
\eqn\ehFd{\eqalign{
\hF
&= (d+ \de)(A+ c) + (A +c)^2, \cr
&= \hF^{2,0} + \hF^{1,1} + \hF^{0,2} \cr
&=   F + \psi + \phi,\cr }
}
where
\eqn\ehFc{\eqalign{
&\hF^{2,0} = F = dA + A^2, \cr
&\hF^{1,1} = \psi = \de A + dc + \{A,c\}, \cr
&\hF^{0,2} = \phi = \de c +c^2, \cr }
}
and satisfies the Bianchi identity
\eqn\ebian{
(d +\de) \hF + [A + c, \hF] = 0.
}
{}From \enpo\ehFd\ehFc, we can get the $\dt$ algebra of \LP\BS
\eqn\det{\eqalign{
\de A &= \psi - dA -\{A,c\},\cr
\de c &= \phi-c^2,\cr
\de \psi &=  -[c,\psi] - d\phi - [A,\phi],\cr
\de \phi &= -[c,\phi].}
}
Note that if we set $\psi=\phi=0$ the $\dt$ algebra\det\ looks like
the ordinary BRS algebra \ebrs\ if we identify $c$ with the Faddev-Popov
ghost and $\de$ with the BRS operator $\db$\K.  The $\dt$ algebra
also contains Witten's topological BRST ($\dw$) algebra
which is given by\W
\eqn\dew{\eqalign{
\dw A = \psi,\qquad\dw\psi=-d\phi-[A,\phi],\qquad\dw \phi = 0.}
}

The condition  $\psi=\phi=0$
is known as the horizontality and, then, eq.\ \ehFd\  reduces to
\eqn\psudo{
(d+ \de)(A+ c) + (A +c)^2 = dA + A^2,
}
which is equivalent to the Russian formula\erusA. Thus, the Russian
formula is nothing but imposing the horizontality condition
in this approach.
It follows that there can  be no analogue  of the Russian formula
in TYM - that is, the full BRST symmetry \det\ of TYM is obtained
if we do not impose the horizontality condition, and instead of
the Russian formula we have non-vanishing extra components of
curvature, which will contribute to the Abelian-anomaly counterpart
in TYM. Consequently, there is no analogue of the non-Abelian anomaly
counterpart in TYM. The Abelian anomaly of TYM absorbs the non-Abelian
anomaly of YM!

However, we will see throughout this paper that the above approach
is not correct.  Note that we did not impose  the extra components
of curvature originated from the transition $(d,\CA)\rightarrow(d+\db,\CA+v)$
to vanish by offhand in the approach described in sect.\ 2, and  the BRS
algebra \ebrs\ and the Russian formula \erusA\ were just byproducts of the
geometrical structures of the Faddev-Popov ghost $v$ and the BRS operator
$\db$. It is also quite strange to say that the extra components $\psi,\phi$
vanish, while $c$ does not vanish identically - unless $c$ is the pure
gauge, because $c$ is the connection one-form which defines the
curvature $\phi$ \ehFc. Thus, we should consider the gauge transformed
connection one-form of $A+c$ as sect.\ 2. to incorporate the correct
BRS structure. Then, we will see that there is an analogue of the
Russian formula in TYM as a natural extension of eq.\ \erusA.
We should not be confused with the notion of the horizontality
condition, which can be stated precisely as `the extra components of
curvature induced by $(\db,v)$ system vanish identically.'
And, it is evident that the structure of BRS symmetry
involving the quantization of Witten's original action of TYM\W,
which has the local gauge symmetry, should be equivalent to the usal
BRS structure of YM\H.

Finally,
there is no need in practice to consider
the gauge transformed connection one-form and the fine points
noted previously as long as the set of Witten's observables is involved,
which is given the invariant polynomial $P(\hF^n)$\K,
This situation is similar to the $2n$-dimensional Abelian anomaly in
the Yang-Mills theory, which is also given by an invariant polynomial
degree $n$. However, the non-Abelian anomaly arises from
the two-high dimensional Abelian anomaly via the {\it descent equation\/},
where the BRS structure of Yang-Mills theory is crucial. Then, discovering
the correct BRS structure will illuminate the counterpart of the
non-Abelian anomaly in TYM theory.

\subsec{Algebra}

In this subsection we reformulate the construction of ref.\K\BBRT\ following
the methods reviewed in sect.\ $2$. A geometrical construction which is
equivalent and somewhat complementary to algebraic one of this subsection is
presented in the appendix A.

Let $A$ and $c$ be the components of some fixed total connection over
$M\times\CU$.  Consider the gauge transformed connection
\eqn\ehAt{
A+c\rightarrow g^{-1} A g + g^{-1}dg +g^{-1} c g + g^{-1}\de g,
}
where we have naturally extends the action of $g$ to $c$.
Let
\eqn\eCACC{\eqalign{
\CA &= g^{-1}A g + g^{-1}dg, \cr
\bCC &= g^{-1}c g + g^{-1}\de g,\cr}
}
such that we replace $A + c$ to $\CA + \bCC$.
Let $\hCF$ denote the transformed total curvature
\eqn\ehCF{\eqalign{
\hCF &= (d + \de)(\CA + \bCC) +(\CA + \bCC)^2 \cr
&= g^{-1}\hF g,\cr
}}
which can be written in the components
\eqn\ehCFc{\eqalign{
&\hCF^{2,0}=\CF  = d\CA + \CA^2, \cr
&\hCF^{1,1}=\Psi = \de \CA+ d\bCC + \{\CA,\bCC\}, \cr
&\hCF^{0,2}=\Phi =\de\bCC + \bCC^2.\cr }
}
The transformed total curvature also satisfies the Bianchi identity
\eqn\eBian{
(d +\de ) \hCF + [\CA + \bCC,\hCF] = 0.
}
{}From \enpo\ehCFc\eBian, we get
\eqn\edeal{\eqalign{
&\de\CA = \Psi - d\bCC - \{\CA,\bCC\}, \cr
&\de\bCC = \Phi- \bCC^2, \cr
&\de\Psi = -[\bCC,\Psi] - d\Phi -[\CA,\Phi], \cr
&\de\Phi = -[\bCC,\Phi],\cr}
}
One can see that the above algebra is {\it formally\/} identical
to $\dt$-algebra in ref.\K. However, there are some important
differences.  Note that we did not start from the orbit space
$\CU/\CG$ but the space of all connection $\CU$. Thus, we can
decompose $\de$ into
\eqn\eded{\eqalign{
\de &= \dt + \db,\cr
\dt^2=\db^2&=\dt\db +\db\dt=0,\cr
}}
and $\bCC$ into
\eqn\ebCCd{\eqalign{\bCC
&= g^{-1}c\,g + g^{-1}\dt g + g^{-1}\db g\cr
&= C + v,\cr}
}
where $v = g^{-1}\db g$ is the Faddev-Popov ghost and $\db$ denote
the exterior derivatives along the gauge group as in sect.\ 2.

Then, from eq.\ \edeal\ or direct computations,
we find the $\db$-algebra
\eqn\edbal{\eqalign{
&\db\CA =  - dv -\{\CA,v\}, \cr
&\db v =  -v^2, \cr
&\db\Psi =  - [v,\Psi] , \cr
&\db \Phi =  -[v,\Phi].\cr
}}
and $\dt$-algebra
\eqn\edtalg{\eqalign{
&\dt \CA = \Psi - dC -\{\CA,C\}, \cr
&\dt C  = \Phi- C^2,  \cr
&\dt \Psi =  -[C, \Psi] - d\Phi -[\CA,\Phi], \cr
&\dt \Phi = -[C, \Phi], \cr
&\dt v = -\db C - \{v,C\}. \cr
}}
Note that the set of equations of eq.\ \edbal\ contains the usual
BRS algebra \ebrs, and they are identical to the extra BRS algebra
introduced by Horne\H\ to fix the remaining local gauge symmetry in Witten's
original action of TYM theory\M. The last term of eq.\ \edtalg\
implies that $\dt$ and $\db$ are not completely decoupled. On the other
hand, one can see in appendix A that the correct variation operator
of the universal bundle is not $\dt$ operator but $\dw$ operator
\eqn\noo{\eqalign{
&\dw \CA  = \Psi, \cr
&\dw \Psi = - d\Phi -[\CA,\Phi], \cr
&\dw \Phi = 0, \cr
&\dw v = 0. \cr
}}
One can see that the ghost
numbers of $(\CA,\Psi,\Phi,v)$ for $\dw$ ($\dt$) algebra are
$(0,1,2,0)$, while $(0,0,0,1)$ in terms of $\db$-algebra - $C$ has
$\dt$ ghost number $1$ and $\db$ ghost number $0$.
Thus we have a natural bigrading structure of the ghost numbers,
which should be preserved independently.

It should be stressed that, if we set $C=0$,
we can recover the usual $\db$ algebra of eq.\ \ebrs\
plus some extra relations
\eqn\ebrss{\eqalign{
&\dt\CA = \Psi,\cr
&\dt\Psi=0,\cr
&\db\Psi =  - [v,\Psi] , \cr
&\dt v =0, \cr
}}
Note that $\dt$ algebra in \edtalg\ can be read as
\eqn\edwalg{\eqalign{
\dt\CA + \{C,\CA\}+dC=\Psi &=\dw\CA,\cr
\dt\Psi +[C,\Psi]=-d\Phi-[A,\Phi] &=\dw\Psi,\cr
\dt\Phi + [C,\Phi] = 0 &= \dw\Phi.\cr
}}
The last two relations imply that the $\dw$ operator is
the covariant derivative
as already noted by \BBRT,
while $\dt$ is the exterior derivative.
The first relation of eq.\ \edwalg\  seems to
be problematic at first sight due to the extra term $dC$.
The reader can see, however, in appendix A  that the
horizontal part of $\dt\CA$ is $\Psi$, which means that $\dw$
operator is the covariant derivative. Note that $\Phi$
vanishes identically for $C=0$, and in this case $\dw$ operator
is equivalent to $\dt$.

Some readers may be confused with our approach, because it was
sometimes believed that $\dt$ algebra reduces to $\dw$ algebra
\dew\ for $C=0$ and reduces to $\db$ algebra \ebrs\ for $\Psi=\Phi=0$.
The difference is that we did not identify $C$ as the Faddev-Popov
ghost. We have shown that for $C=0$ the $\de$ operator decouple
completely into $\db$ plus $\dt$ operators, and the later reduces to $\dw$
algebra with $\Phi =0$. It is obvious that a covariant derivative
is equivalent to ordinary one when a gauge orbit of a connection form
vanishes. Note that there is an additional term $\db\Psi=-[v,\Psi]$
in our $\db$ algebra \ebrss\ for $C=0$ unlike the original $\db$ algebra
of \ebrs. It is, however, unnecessary and indeed not correct to set
$\dt\CA =\Psi=0$ to recover the usual BRS algebra, because there
is no reason to set $\Psi$ (tangent to a local cross section of
$\CU/\CG\rightarrow \CU$) to zero. We just do not use the extra
component $\Psi$ and its  BRS variation in the BRS quantization of the
Yang-Mills theory. More detailed analysis can be found in appendix A.

Finally, we remind the readers that Horne's
approach is correct method in the quantization of TYM\H.
In the next subsection, we will see that the real utility
of $\dt$ algebra is its ability of local trivialization.

\subsec{Russian formula and descent equation}

Note that
\eqn\eRusa{\eqalign{\hCF
&= (d+\dt+\db)(\CA + C + v)+(\CA + C +v)^2 \cr
&= (d + \dt)(\CA + C) +(\CA+C)^2,\cr}
}
where we have used eqs.\ \edbal\edtalg.  One can see that the above
equation is a generalization of the {\it Russian formula} \erusA.
In the limit of $C=0$ it is not entirely identical to
eq.\ \erusA;
\eqn\eRusaa{
(d+\dt+\db)(\CA + v)+(\CA +v)^2= d\CA + \CA^2 +\dt\CA =\CF +\Psi,
}
however, after throwing away $\dt$ we get desired result\foot{
In preparing the revised version of this manuscript
I have found, rather suprisingly, that a similar formula for eq.\ \eRusaa\
is already appeared in ref. \ZuminoC. The BRST symmetry of TYM had
been almost discovered!}.

{}From the Bianchi identity \eBian
\eqn\eBianc{
(d + \dt +\db)\hCF + [\CA + C + v, \hCF] = 0,
}
we can define an symmetric invariant polynomial
of degree $n$, $P(\hCF^n)$;
\eqn\esip{
(d + \dt + \db)P(\hCF^n) = 0.
}
One can also find the another Bianchi identity
\eqn\eBianca{
(d + \dt)\hCF + [\CA + C, \hCF] = 0,
}
which leads
\eqn\esipa{
(d + \dt)P(\hCF^n) = 0.
}

Note that $P(\hCF^n)$ is identical to $P(\hF^n)$.
A simple consequence of the above identity \esipa\
is so called the {\it topological descent equation}\K.
That is, if we expand $P(\hCF^n)$ in powers of the $\dt$ ghost
number
such that
\eqn\epvt{P(\hCF^n)
=\tCW_{2n}{}^{0,0}+\tCW_{2n-1}{}^{0,1}+\cdots+\tCW_{0}{}^{0,2n},
}
where the superscripts indicate $\db$ and $\dt$ ghost numbers,
respectively, and the subscript indicate the space-time form
degree, eq.\ \esipa\ leads the {\it topological descent equation}
\eqn\esdeca{\eqalign{
d\tCW_{2n}{}^{0,0} &= 0, \cr
d\tCW_{2n-1}{}^{0,1} + \dt\tCW_{2n}{}^{0,0} &= 0, \cr
d\tCW_{2n-2}{}^{0,2} + \dt\tCW_{2n-1}{}^{0,1} &= 0, \cr
\vdots& \cr
d\tCW_{0}{}^{0,2n} + \dt\tCW_{1}{}^{0,2n-1}&= 0, \cr
\dt\tCW_{0}{}^{0,2n} &= 0.\cr}
}
It is well-known that the integration of
$\tCW_{2n-\ell}{}^{0,\ell}$
in \epvt\ over  $2n-\ell$ cycle $\ga_{{}_{2n-\ell}}$ of $M$ is the
Witten's observable\foot{Note that our convention is the
somewhat different from the usual one in denoting a Witten's observable.
In  Donaldson-Witten theory the dimension of $M$ is four, $0\leq
\hbox{dim}(Y)\leq 4$ and there is no restriction such that we should only
consider second-rank invariant polynomial, as everybody knows.
Thus, if we restrict $M$ is four dimensional, $P(\hCF^n)$ can be
regarded as an element of $H^{2n}(M\times\CU/\CG)$.
Then a Witten's observable $\tW_{2n-\ell}{}^{0,\ell}$ should reads as
$\tW_{\ell^\prime}{}^{0,2n-\ell^\prime}$, where $0\leq\ell^\prime\leq 4$,
which is an element of $H^{2n-\ell^\prime}(\CU/\CG)$. Similarly, the
cycle $\ga_{{}_{2n-\ell}}$ should reads as $\ga_{{}_{\ell^\prime}}$.
I hope the readers may not be confused with our conventions.}
for Donaldson's polynomial invariant\W\K;
\eqn\ewo{
\int_{\ga_{{}_{2n-\ell}}}\tCW_{2n-\ell}{}^{0,\ell}
\equiv \tW_{2n-\ell}{}^{0,\ell}.
}
{}From the topological descent equation \esdeca\ we can see that
the Witten's observable
$\tW_{2n-\ell}{}^{0,\ell}$, ($k=0,\ldots,2n$)
is $\dt$ closed
\eqn\ecdi{
\dt\int_{\ga_{{}_{2n-\ell}}}\tCW_{2n-\ell}{}^{0,\ell}
=\dt \tW_{2n-\ell}{}^{0,\ell} =0,
}
as well as $\db$ closed, which follows from eq.\ \esip\esipa.

However, this is not the end of the story. Eq. \esip\esipa\
imply that we have an identity by the Poincar\'{e} lemma
\eqn\eRusb{\eqalign{
P(\hCF^n)
&=(d + \dt + \db)\CW_{2n-1}(\CA + C + v,\hCF)\cr
&= (d + \dt )\CW_{2n-1}{}^0 (\CA + C, \hCF),\cr}
}
where $\CW_{2n -1}(\CA+C+v,\hCF)$ denotes the extended
Chern-Simons form. It should be noted that the above formula
valid only after {\it local trivializtion\/} and does not imply
the Witten's observables are trivial\M. Note also that
we can not replace $\dt$ with $\dw$ in eq.\ \eRusb,
while we can do for eq.\ \esdeca\ecdi.
Expanding $\CW_{2n-1}(\CA + C + v,\hCF)$
with powers of $v$
\eqn\eepv{
\CW_{2n-1}(\CA + C + v,\hCF) = \CW_{2n-1}{}^0(\CA + C,\hCF)
+ \CW_{2n-2}{}^1 + \cdots +\CW_{0}{}^{2n-1},
}
where the superscript indicates the power of $v$ (the $\db$ ghost
number) and the subscript indicates the space-time form degree
plus the $\dt$ ghost number.  Thus we can get an descent
equations which have the same form and origin with
eq.\ \edec
\eqn\eedec{\eqalign{
(d+\dt)\CW_{2n-2}{}^1(\CA+C,v,\hCF)
+\db\CW_{2n-1}{}^0(\CA+C,\hCF)&=0,\cr
(d+\dt)\CW_{2n-3}{}^2(\CA+C,v,\hCF)
+\db\CW_{2n-2}{}^1(\CA+C,v,\hCF)&=0,\cr
\vdots& \cr
(d+\dt)\CW_{0}{}^{2n-1}(v)
+\db\CW_{1}{}^{2n-2}(\CA+C,v)&=0,\cr
\db\CW_{0}{}^{2n-1}(v) &= 0.\cr}
}
We shall call the above relations  {\it extended descent
equations}.

We can expand the each relations of \eedec\ in powers of $\dt$ ghost
number.
For an example, consider the first relation of \eedec\
$$
(d+\dt)\CW_{2n-2}{}^1(\CA+C,v,\hCF)
+\db\CW_{2n-1}{}^0(\CA+C,\hCF)=0,
$$
which leads
\eqn\eee{\eqalign{
d\CW_{2n-2}{}^{1,0} +\db\CW_{2n-1}{}^{0,0}&=0,\cr
d\CW_{2n-3}{}^{1,1}+\dt\CW_{2n-2}{}^{1,0}+\db\CW_{2n-2}{}^{0,1}&=0,\cr
\vdots& \cr
d\CW_{0}{}^{1,2n-2}+\dt\CW_{1}{}^{1,2n-3}+\db\CW_{1}{}^{0,2n-2}&= 0, \cr
\dt\CW_{0}^{1,2n-2}+\db\CW_{0}{}^{0,2n-1}&= 0,\cr}
}
after using the expansions in terms of $\dt$ ghost number
\eqn\expansion{
\CW_{2n-1-k}{}^{k} = \sum_{k^\prime=0}^{2n-1-k}
\CW_{2n-1-k-k^\prime}{}^{k,k^\prime},
}
%$$
%\CW_{2n-1}{}^0
%= \CW_{2n-1}{}^{0,0} + \CW_{2n-2}{}^{0,1} + \cdots + \CW_{0}{}^{0,2n-1},
%$$
%$$
%\CW_{2n-2}{}^1
%= \CW_{2n-2}{}^{1,0} + \CW_{2n-3}{}^{1,1} + \cdots + \CW_{0}{}^{1,2n-2}.
%$$
such that the superscripts $k$ and $k^\prime$ of
$\CW_{2n-1-k-k^\prime}{}^{k,k^\prime}$
denote $\db$ and $\dt$ ghost numbers,
respectively, and the subscript $2n-1-k-k^\prime$ indicate the space-time
form degree.
Similarly, the second relation of \eedec
\eqn\eedecb{
(d + \dt)\CW_{2n-3}{}^2(\CA + C,v, \hCF) +
\db\CW_{2n-2}{}^1(\CA + C, \hCF) = 0,
}
leads
\eqn\ess{\eqalign{
d\CW_{2n-3}{}^{2,0} +\db\CW_{2n-2}{}^{1,0}&=0,\cr
d\CW_{2n-4}{}^{2,1}+\dt\CW_{2n-3}{}^{2,0}+\db\CW_{2n-3}{}^{1,1}&=0,\cr
d\CW_{2n-5}{}^{2,2}+\dt\CW_{2n-4}{}^{2,1}+\db\CW_{2n-4}{}^{1,2}&=0,\cr
\vdots& \cr
d\CW_{0}{}^{2,2n-3}+\dt\CW_{1}{}^{2,2n-4}+\db\CW_{1}{}^{1,2n-3}&= 0, \cr
\dt\CW_{0}^{2,2n-3}+\db\CW_{0}{}^{1,2n-2}&= 0,\cr}
}

Note that the first relations of \eee\ and \ess\ are identical to the
first and second relations of eq.\ \edec\ respectively. Similarly, if we
expand the remaining relations of \eedec, the first relation
of each of the sub-descent equations reduces to the relations in
the descent equation of Yang-Mills \edec\ subsequently.
Note also, for an example, that the last relations of \eee,
\ess\ and the other sub-descent equations of \eedec\ lead
\eqn\efn{\eqalign{
\dt\CW_{0}^{1,2n-2}+\db\CW_{0}{}^{0,2n-1}&= 0,\cr
\dt\CW_{0}^{2,2n-3}+\db\CW_{0}{}^{1,2n-2}&= 0,\cr
                                         &\vdots\cr
\dt\CW_{0}^{2n-1,0}+\db\CW_{0}{}^{2n-2,1}&= 0,\cr
\db\CW_{0}{}^{2n-1,0}&= 0.\cr}
}
Clearly, the above descent equation is originated from a part of the
extended Russian formula \eRusa
\eqn\epdec{
\Phi = \dt C + C^2 = (\dt + \db)(C + v) +(C+v)^2,
}
and from a part $P(\Phi^n)=\tCW_{0}{}^{0,2n}$  of the invariant polynomial
$P(\hCF^n)$ after the similar procedure discussed in sect.\ 2.

Each term of the fully expanded descent equations can be exactly calculated
by the following procedures. Note that the structure of the extended
descent equation \eedec\ is formally identical to that of the Yang-Mills
case \edec\ if we set $\hd = d + \dt$ and $\hCA = \CA + C$.  Thus we can
follow the fully developed and well-known methods proposed in
\ZuminoA\ZuminoB\MSZ\ZWZ\ to get $\CW_{2n-1-k}{}^k$. Then, by expanding
$\CW_{2n-1-k}{}^{k}$ in terms of the $\dt$ ghost number, we can get
$\CW_{2n-1-k-k^\prime}{}^{k,k^\prime}$.
Following the method developed in ref.\ZuminoB,
we describe the detailed procedures and some explicit results of
calculation in the appendix B.

\newsec{Non-Abelian Anomaly Counterpart in TYM}

In the previous section we have extended the Russian formula and the
descent equation, and we have seen that the extended descent equation
contains the Yang-Mills descent equation \edec\ as a sub-relation. In
particular, eq.\ \eedecb\ after the expansion in terms of $\dt$-ghost
number leads the Wess-Zumino consistency condition as a sub-relation,
i.e.\ the first relation of eq.\ \ess
\eqn\ecos{
d\CW_{2n-3}{}^{2,0} +\db\CW_{2n-2}{}^{1,0} = 0,
}
which is identical to \ewz.

Then, it is natural to ask whether $\CW_{2n-2}{}^1$ in general is
the non-Abelian anomaly counterpart in topological Yang-Mills theory and
whether eq.\ \eedecb\ is the corresponding consistency condition. Clearly
$\CW_{2n-2}{}^1$ is also linear in $v$ as $\z_{2n-2}{}^1$
($\CW_{2n-2}{}^{1,0}$), and if we integrate \eedecb\ by a $2n-2$
dimensional cycle $\tilde\ga_{{}_{2n-2}}$ of the product space
$M\times \CU/\CG$ we get
\eqn\eecos{
\db\int_{\tilde\ga_{{}_{2n-2}}} \CW_{2n-2}{}^1 = 0,
}
which can be interpreted as a consistency condition.

To be more precise, consider integration the $i$th relation of eq.\ \ess\
over a $2n-1-i$ cycle $\ga_{{}_{2n-1-i}}$ of $M$
\eqn\eint{\eqalign{
\db W_{2n-2}{}^{1,0} &=0,\cr
\dt W_{2n-2-\ell}{}^{2,\ell-1} + \db W_{2n-2-\ell}{}^{1,\ell} &=0,
\quad\hbox{for}\quad \ell= 1,\ldots, 2n-2 \cr}
}
where we generally denote
\eqn\egdt{
W_{2n-1-k-k^\prime}{}^{k,k^\prime} = \int_{\ga_{{}_{2n-1-k-k^\prime}}}
\!\!\!\!\!\CW_{2n-1-k-k^\prime}{}^{k,k^\prime},
}
where $\ga_{{}_{2n-1-k-k^\prime}}$ is a $2n-1-k-k^\prime$ dimensional
cycle of $M$.
%$$
%W_{2n-2-\ell}{}^{1,\ell} = \int_{\ga_{{}_{2n-2-\ell}}}\!\!\!\!\!
%\CW_{2n-2-\ell}{}^{1,\ell}, \quad
%W_{2n-2-\ell}{}^{2,\ell-1} = \int_{\ga_{{}_{2n-2-\ell}}}\!\!\!\!\!
%\CW_{2n-2-\ell}{}^{2,\ell-1}.
%$$
Note that $W_{2n-2-\ell}{}^{1,\ell}$ is a zero form over $M$ and
a $\ell$-form over $\CU/\CG$, respectively, and $\dt$ is
the exterior derivative over $\CU/\CG$. Thus, if we integrate \eint\
over a $\ell$-dimensional cycle $\ga^{\prime}_{{}_\ell}$ of $\CU/\CG$,
we get
\eqn\eintt{
\db\int_{\ga^{\prime}_{{}_\ell}} W_{2n-2-\ell}{}^{1,\ell} = 0.
}
That is, the resulting
$\int_{\ga^{\prime}_{{}_\ell}} W_{2n-2-\ell}{}^{1,\ell}$
is also linear in $v$ and satisfies the consistency condition.

Integration of eq.\ \eee\ over appropriate cycle of $M$
leads
\eqn\ecli{
\dt W_{2n-2-j}{}^{1,j} = -\db W_{2n-2-j}{}^{0,j+1},\quad\hbox{for }
j=0,\dots, 2n-2
}
Thus we can see that $W_{2n-2-j}{}^{1,j}$ is $\db$ ($\dt$) closed up to $\dt$
($\db$) exact term.  If we integrate the both sides of \ecli\ by
a $j+1$ dimensional cycle $\ga^\prime_{{}_{j+1}}$ of $\CU/\CG$, we can see
that
\eqn\eccli{
\db\int_{\ga^{\prime}_{{}_{j+1}}} W_{2n-2-j}{}^{0,j+1} = 0.
}
We can repeat the same procedures for the remaining sub-descent equations
of eq.\ \eedec\ after using the expansion \expansion\ and conclude that
$\int_{\ga^{\prime}_{{}_{k^\prime}}} W_{2n-1-k-k^\prime}{}^{k,k^\prime}$
is $\db$ invariant
\eqn\eclos{
\db \int_{\ga^{\prime}_{{}_{k^\prime}}}
W_{2n-1-k-k^\prime}{}^{k,k^\prime} =0.
}

We can see that the $\dt$ with $\db$ cohomology class of
$W_{2n-1-k-k^\prime}{}^{k,k^\prime}$ depends only on
the homology class of $\ga$.
That is, if $\ga_{{}_{2n-1-k-k^\prime}}$ is a boundary,
say $\ga_{{}_{2n-1-k-k^\prime}}=\rd\be_{{}_{2n-k-k^\prime}}$, then
\eqn\ehomo{\eqalign{
W_{2n-1-k-k^\prime}{}^{k,k^\prime}
&= \int_{\ga_{{}_{2n-1-k-k^\prime}}} \CW_{2n-1-k-k^\prime}{}^{k,k^\prime}\cr
&= \int_{\be_{{}_{2n-k-k^\prime}}} d\CW_{2n-1-k-k^\prime}{}^{k,k^\prime}\cr
&= -\dt\int_{\be_{{}_{2n-k-k^\prime}}}\CW_{2n-k-k^\prime}{}^{k,k^\prime -1}
- \db\int_{\be_{{}_{2n-k-k^\prime}}}\CW_{2n-k-k^\prime}{}^{k-1,k^\prime}.\cr
}}
It is well known that the $\dt$ cohomology class (or $\dw$ cohomology class)
of Witten's observables
$\tW_{2n-\ell}{}^{0,\ell}$ depends only on the homology class of cycle in $M$
\eqn\whomo{\eqalign{
\tW_{2n-\ell}{}^{0,\ell}
=\int_{\ga_{{}_{2n-\ell}}}\tCW_{2n-\ell}{}^{0,\ell}
&=\int_{\be_{{}_{2n+1-\ell}}}d\tCW_{2n-\ell}{}^{0,\ell}\cr
&=-\dt\int_{\be_{{}_{2n+1-\ell}}}\tCW_{2n+1-\ell}{}^{0,\ell-1}\cr
}}
where $\ga_{{}_{2n-\ell}} = \rd \be_{{}_{2n+1-\ell}}$.
If we further integrate the both sides of eq.\ \ehomo\ over
a $k^\prime$ dimensional cycle $\ga^\prime_{{}_{k^\prime}}$ of $\CU/\CG$
\eqn\jack{
\int_{\ga^\prime_{{}_{k^\prime}}}\int_{\ga_{{}_{2n-1-k-k^\prime}}}
\CW_{2n-1-k-k^\prime}{}^{k,k^\prime}
=-\db\int_{\ga^\prime_{{}_{k^\prime}}}\int_{\be_{{}_{2n-k-k^\prime}}}
\CW_{2n-k-k^\prime}{}^{k-1,k^\prime},
}
which show that
the $\db$ cohomology class of
$\int_{\ga^\prime_{{}_{k^\prime}}}W_{2n-1-k-k^\prime}{}^{k,k^\prime}$
depends only on homology class of $\ga$.
Note also that, if $\ga^\prime$ is a boundary of $\be^\prime$,
say  $\ga^\prime_{{}_{k^\prime}}=\rd\be^\prime_{{}_{k^\prime +1}}$, we
have
\eqn\jake{\eqalign{
\int_{\ga^\prime_{{}_{k^\prime}}}W_{2n-1-k-k^\prime}{}^{k,k^\prime}
&=-\db\int_{\be^\prime_{{}_{k^\prime+1}}}\dt
W_{2n-k-k^\prime}{}^{k-1,k^\prime}\cr
&=\db\int_{\be^\prime_{{}_{k^\prime+1}}}\db\,
W_{2n-k-k^\prime}{}^{k-2,k^\prime+1}\cr
&=0,\cr}
}
which show that
$\int_{\ga^\prime_{{}_{k^\prime}}} \int_{\ga_{{}_{2n-1-k-k^\prime}}}
\!\!\!\!\!\CW_{2n-1-k-k^\prime}{}^{k,k^\prime}$
vanishes if $\ga^\prime$ is trivial in homology.

Up to now we have developed an analogy with the local Yang-Mills theory
and  introduced the non-Abelian anomaly counterpart of TYM.
We have shown that there are large
classes of $\db$ invariant quantities
which $\db$ cohomology class only depends on  homology
class of $\ga$ .
These properties should be compared with the Witten's observables, which
are $\dt$ invariant as well as $\db$ invariant and their $\dt$ cohomology
classes are depends only on the homology class of $\ga$.

It should be
stressed that the notions of cohomology associated with a Witten's
observable $\tW_{2n-k}{}^{0,k}$ and
with $\int_{\ga^\prime_{{}_{k^\prime}}}W_{2n-1-k-k^\prime}{}^{k,k^\prime}$
(as well as $W_{2n-1-k-k^\prime}{}^{k,k^\prime}$) are indeed different.
The later is the integrals of a locally defined
density and its cohomology is so called the {\it local
cohomology}\ZuminoC, and we can not interchange $\dt$ cohomology with
$\dw$ cohomology unlike the former.

One can easily show that it is {\it non-trivial\/} in the sense of the local
cohomology.
The non-triviality of $\CW_{2n-1-k}{}^{k,0}$ in the sense of
local cohomology has been prove in sect. 3 of ref.\MZ.  Following the same
procedures after replacing $(d,\CA)$ to $(d +\dt, \CA + C)$, one can easily
prove that one can not write
\eqn\entivi{
\CW_{2n-1-k}{}^k = (d +\dt){\hat{\eta}}_{2n-2-k}{}^k
+ \db{\hat{\theta}}_{2n-1-k}{}^{k-1},
}
where $\hat{\eta},\hat{\theta}$ are some local expressions.
Furthermore,
a mathematical expression for the triviality of
$\CW_{2n-1-k-k^\prime}{}^{k,k^\prime}$
\eqn\etrivc{
\CW_{2n-1-k-k^\prime}{}^{k,k^\prime} =
d \eta_{2n-2-k-k^\prime}{}^{k,k^\prime}
+\dt\xi_{2n-1-k-k^\prime}{}^{k,k^\prime -1}
+ \db\theta_{2n-1-k-k^\prime}{}^{k-1,k^\prime},
}
can be viewed as an expansion of eq.\ \entivi, that is impossible.

\newsec{Conclusion and Further Study}

At present, there are two important open problems in Donaldson-Witten
theory. One is to
extend the theory beyond the {\it stable region}, and the
other is to find some consistent topological symmetry breaking mechanism.
It is well known that there are many serious problems beyond the stable
region, i.e.\  we can not avoid the reducible connections which
contribute to the singularities and the non-compactness of the
space of $\dw$ fixed points $\tilde{\CM}$ - the instanton moduli space
$\CM$ with the solution space of $D\Phi=0$.

One of the main motivations of this paper is that TFT's should be
a generalizations of the local theories.  Clearly
TFT's have no local degrees of freedom that unless the topological
symmetry is broken down to the local symmetry, there seems to be no way
to describe the local theories.
However, we find that the algebraic structure of TYM theory is rich enough
to contains that of local Yang-Mills theory, such that the Yang-Mills
theory can be regarded as a local sector of TYM theory.
In particular, we
have suggested the Russian formula, descent equation and non-Abelian
anomaly counterparts in TYM,
which are natural extensions of those
of local Yang-Mills theory.

It remains an open question to know the precise mathematical and physical
meanings of the non-Abelian anomaly counterpart in TYM.
There is at least one indication
of the possible physical application of the non-Abelian anomaly.
Note that the zero modes of Faddev-Popov ghost $v$ will arise
due to the reducible connections or due to the Gribov ambiguity\Gri\S.
Then, we have the net violation
of $\db$ ghost number as well as that of $\dw$ ghost number
zero-modes. Thus, some appropriate set of observables should be
inserted to the correlation function of TYM to absorb the both
kinds of zero modes. Because the Witten's observables have no
$\db$ ghost number, we need other set of observables which have
non-zero $\db$ ghost number. Such an observable should be $\dw$
as well as $\db$ closed and non-trivial in the sense of global
topology. If there is no such an observable,
not only the topological interpretation of correlation function
becomes impossible but also the correlation function itself can not be
well defined. A way out of this problem is to include $\dw$
non-invariants, but preserving $\db$ invariance, into the correlation
function. Then, a natural candidate may be the consistent anomaly
counterpart in TYM
\eqn\eosb{
\int_{\ga_{{}_{\ell}}}\int_{\ga^\prime_{{}_{2n-\ell}}}
\CW_{\ell}{}^{1,2n-\ell},
}
which  is $\db$ invariant and its  $\db$ cohomology ({\it local cohomology})
class depends only on the homology class of $\ga$.

To be definite, let $\triangle U$ and $\triangle u$ denotes the net
violations of $\dt$ and $\db$ ghost numbers of zero-modes, respectively.
Then the correlation function
\eqn\eend{
\left<\prod_{i=1}^{r}\tW_{k_i}{}^{0,2n-k_i}
\prod_{j=1}^{\triangle
u}\int_{\ga^\prime_{2n-\ell_j}}W_{\ell_j}{}^{1,2n-\ell_j}
\right>,
}
may be well defined for
$$
\triangle U = \sum_{i=1}^{r}(2n- k_i).
$$
In the semi-classical limit the above correlation function reduce
to an integration over the space of fixed points of $\dw$ and $\db$
symmetries,
and the integrand is certain cohomology class on the fixed points,
which can be obtained by replacing the fields of the inserted
quantities by their zero-modes. Then, one can immediately see that
the above correlation function is factorized
\eqn\endd{
\left<\prod_{i=1}^{r}\tW_{k_i}{}^{0,2n-k_i}\right>_{\dw X=0}
\left<\prod_{j=1}^{\triangle
u}\int_{\ga^\prime_{2n-\ell_j}}W_{\ell_j}{}^{1,2n-\ell_j}
\right>_{\db\CA =0},
}
where $<\cdots>_{\dw X=0}$ denotes the integration over $\dw$ fixed
points, which is identical to the original correlation function of TYM.
The additional terms $<\cdots>_{\db\CA=0}$
is then the integration the local $\db$ cohomology class over the
space of $v$ zero modes. I do not know what kind of topological
meaning, if any, it may have.

Note, however, that if there are zero modes due to the reducible
connections, the validity of semi-classical approximation becomes
doubtful, the factorization of \endd\ becomes doubtful and the
topological interpretation of the correlation function becomes
unclear\WW.
Note also that it is quite unpleasant to observe that the consistent
anomaly involve the field $C$ in general, which never appears
explicitly in the completely fixed action of TYM. It will be
also practically impossible to define non-trivial homology
cycle in the orbit space $\CU/\CG$ or in the instanton moduli space
unless the connection is generic. Thus, the use of the consistency
anomaly seems to be restricted to the stable region.

\ack
I am grateful to J.M.\ Park for reading on the manuscript.
\vfill
\eject

\appendix{A}{Geometrical Understanding of BRST Symmetries}

In this appendix we discuss the geometrical origin of the various BRST
algebras - $\db$, $\dt$ and $\dw$) algebras
based on the universal bundle formalism\AS.
The presentations of this section are  motivated by the appendix of
the ref.\Stora, and we generally follow the conventions of ref.\ZuminoA.

Consider a principal $G$-bundle $P$ over base
space $M$. Let $\CU$ denote the affine space of all connections
on $P$ and $\CG$ denote bundle automorphism, which acts as the
gauge symmetry group, leaving a base point fixed.
Now consider a principal $\CG$-bundle over base space $(P\times
\CU/\CG)$ where $G$ acts freely; $\left(P\times\CU,
\CG, (P\times\CU)/\CG\right)$.  The base space of the
above bundle itself can be regarded as a principal $G$-bundle
over $M\times\CU/\CG$, which is called the {\it universal
bundle}\AS
\eqn\ub{((P\times \CU)/\CG,G,M\times \CU/\CG).
}
It is convenient to start from a pull backed bundle $Q$ over
$M\times \CU$ from the universal bundle \ub. One can locally
parametrize $\CU$ by
\eqn\ca{
\CA = g^{-1}A\,g + g^{-1}dg,
}
where $A$ denotes a fixed connection one-form tangent to $M$,
and $g$ is an element of gauge group
$\CG$ which depends  on the space-time coordinates $x^\mu$,
some group parameter$\la^i$.

An arbitrary variation on $\CA$ is
\eqn\deca{
\de \CA = g^{-1} \de A\, g - d_{\CA}(g^{-1}\de g),
}
where we will interpret the operator $\de$ as the exterior
derivative over $\CU$.

If we decompose the operator $\de$
as
\eqn\dec{
\de = \dt + \db,
}
such that $\db$ is the variation (exterior derivation) along
gauge group $\CG$\foot{
Since $\CU$ itself can be viewed
as an principal bundle $\CU\rightarrow \CU/\CG$ over $\CU/\CG$,
$g$ depends also on coordinates of $\CU/\CG$ via a local section.
That is, $\dt g$ does not vanish in general.}
\eqn\edb{
\db = d\la^i\rF{\la^i},
}
and $\dt$ is the exterior derivative over the orbit space $\CUG$
\eqn\enil{
\de^2 =\dt^2 =\db^2 = \dt\db +\db\dt = 0.
}
Then \deca\ becomes
\eqn\decad{
\de\CA = g^{-1}\dt A\,g -d_\CA(g^{-1}\dt g) - d_\CA
(g^{-1}\db g).
}
such that
\eqn\decada{\eqalign{
\dt\CA &= g^{-1}\dt A\, g - d_{\CA}( g^{-1}\dt g),\cr
\db\CA &= - d_\CA (g^{-1}\db g),\cr}
}
where we have used $\db A=0$.

Introducing the connection one-form on $\CU$
\eqn\cc{-G_{\CA}d^{*}_{\CA}\de\CA\equiv \CC,
}
where
$$
G_{\CA} =\left( d^{*}_{\CA} d_{\CA}\right)^{-1}.
$$
The connection one-form $\CC$ can be also decomposed as
\eqn\cd{\CC = -G_{\CA}d^{*}_{\CA}\de\CA
= - G_{\CA}d^{*}_{\CA}\dt\CA - G_{\CA}d^{*}_{\CA}\db\CA,
}
Then one can define the Faddev-Popov ghost $v$ as
\eqn\cg{
- G_{\CA}d^{*}_{\CA}\db\CA = g^{-1} \db g \equiv v,
}
which is the connection one-form along $\CG$, and the BRS algebra
naturally follows
\eqn\brs{\eqalign{
\db\CA &=- dv - \CA v -v\CA \equiv - d_{\CA}v,\cr
\db v &= - v^2.\cr}
}
The total connection one-form over $M\times \CU$ is
\eqn\tc{
\CA + \CC =\CA - G_{\CA} d_{\CA}^{*} \de\CA,
}
and total curvature over $M\times\CU$ is
\eqn\tf{\eqalign{\hCF
= &(d + \de)(\CA - G_{\CA}d_{\CA}^{*}\de\CA)
+\left(\CA-G_{\CA}d_{\CA}^{*}\de\CA\right)^2\cr
= &\CF+ \left(1-d_\CA G_\CA d^*_\CA\right)\de\CA
- \de\left(G_\CA d^*_\CA\de\CA\right)
+\left(G_\CA d^*_\CA\de\CA\right)^2,\cr }
}
which can be written in components
\eqn\tfc{\eqalign{
&\hCF^{2,0}\equiv\CF=d\CA+\CA^2,\cr
&\hCF^{1,1}=\left(1-d_\CA G_\CA d^*_\CA\right)\de\CA,\cr
&\hCF^{0,2}=-\de\left(G_\CA d^*_\CA\de\CA\right)
+\left(G_\CA d^*_\CA\de\CA\right)^2.\cr }
}
Using the decomposition \decada\ and \decada\cd\cg\brs\ we can get
\eqn\foo{\eqalign{\hCF^{1,1}
& = \left(1-d_\CA G_\CA d^*_\CA\right)(\dt\CA+\db\CA),\cr
& = \left(1-d_\CA G_\CA d^*_\CA\right)\dt\CA,\cr}
}
and
\eqn\fzt{\eqalign{ \hCF^{0,2}
=&-\dt\left(G_\CA d^*_\CA\dt\CA\right)
+\left(G_\CA d^*_\CA\dt\CA\right)^2 +\db v + v^2\cr
&+\dt v -\db(G_\CA d^*_\CA\dt\CA) -\{G_\CA d^*_\CA\dt\CA,v\}\cr
=&-\dt\left(G_\CA d^*_\CA\dt\CA\right)
+\left(G_\CA d^*_\CA\dt\CA\right)^2.\cr}
}
Thus, $\hCF$ is also given by
\eqn\rus{\hCF
= \CF + \left(1-d_\CA G_\CA d^*_\CA\right)\dt A
- \dt\left(G_\CA d^*_\CA\dt\CA\right)
+\left(G_\CA d^*_\CA\dt\CA\right)^2,
}
which means $\hCF$ is the total curvature over $M\times\CU/\CG$.
The identities \tf\rus\ an {\it extended Russian formula\/} \eRusa.
That is, if one restricts the variation of $\CA$ in \deca\ to the gauge
group direction\Stora, the above two equation \tf\rus\ lead to
the well-known {\it Russian formula\/} \erusA.

Note that $\left(1-d_\CA G_\CA d^*_\CA\right)$ is the horizontal
projection.  Then
\eqn\hpo{
\hCF^{1,1} = \left(1-d_\CA G_\CA d^*_\CA\right)\dt\CA \equiv \de^H\!\CA,
}
where $\de^H$ denotes the operator for horizontal variation.
Furthermore, direct calculation  shows that
\eqn\hpop{
\de^H(\de^H\!\CA) = - d_\CA\hCF^{0,2},\qquad \de^H\,\hCF^{0,2} = 0.
}
Being the horizontal variation, $\de^H\!\CA$ should satisfy
\eqn\hc{d^*_\CA(\de^H\!\CA) = 0.}
Applying $\de^H$ to the above condition,  we can get
\eqn\hcc{\eqalign{
\de^H\!\left(d^*_\CA\de^H\!\CA\right)
&= [\de^H\!*\!\CA,\de^H\!\CA] - d^*_\CA(\de^H(\de^H\!\CA))  \cr
&=[\de^H\!*\!\CA,\de^H\!\CA] + d^*_\CA d_\CA\Phi \cr
&= 0,\cr}
}
which can be read as
\eqn\Fzt{\Phi=\hCF^{0,2} = -G_\CA[\de^H\!*\!\CA,\de^H\!\CA].}
Thus we have obtained Atiyah-Singer's results\AS.

Note that if we denote $\hCF^{1,1}\equiv\Psi$, $\hCF^{0,2}=\Phi$
such that
\eqn\jack{
\hCF = \CF + \Psi + \Phi,
}
and $\de^H\equiv\dw$, we can get  Witten's BRST algebra
\eqn\edw{
\dw\CA = \Psi,\qquad \dw\Psi = - d_\CA\Phi,\qquad\dw\Phi=0.
}
Let
\eqn\eC{
C \equiv -G_\CA d^*_\CA\dt\CA,
}
such that \cd\ becomes
\eqn\eCC{
\CC = C + v.}
Then \foo\fzt\ lead to the $\dt$ algebra\BS
\eqn\edt{\eqalign{
&\dt\CA = \Psi -d_\CA C,\cr
&\dt C = \Phi -C^2,\cr
&\dt\Psi = -[C,\Psi] - d_\CA\Phi,\cr
&\dt\Phi = -[C,\Phi],\cr }
}
with
\eqn\nunn{
\dt v=-\db C - \{C,v\}.
}

Using
\eqn\dtp{\dt\CA = \Psi - d_\CA C,
}
one can find that
\eqn\dtpc{
-G_\CA d^*_\CA\dt\CA =-G_\CA d^*_\CA\Psi + C.
}
Then, the following condition is crucial
in the self-consistency of $\dt$ algebra
\eqn\epsit{
d^*_\CA \Psi = 0,}
which is identical to \hc.
Note also that
\eqn\more{
\hCF^{1,1} = \left(1-d_\CA G_\CA d^*_\CA\right)(\dt\CA+\db\CA)
=\Psi\equiv \de^H\!\CA.}
Thus, we can see that the horizontal part\foot{Eq. \hpo\ implies that
$\dw$ is the exterior covariant derivative for $\dt$, while eq.\ \more\
implies that $\dw$ is the exterior covariant derivative for $\de =\dt+\db$.
There is no conflict between the two interpretations due to the extended
Russian formula \eRusa\tf\rus. It is enough to examine an identity
$$\dt \Phi +[C,\Phi]=(\dt + \db)\Phi + [C+v,\Phi]=\dw\Phi=0$$.}
of $\de\CA$ is $\Psi$
and the vertical part of $\de\CA$ is $-d_\CA C -d_\CA v$.
Note that the vertical part of $\de\CA$ is not just the BRS
variation of $\CA$. Precisely speaking, the BRS variation of
$\CA$ is the vertical part of $\de\CA$ if we restrict $\de$ to
$\db$  or the component of the vertical part which has $\db$
ghost number one.
Clearly, for $C=0$ the vertical part
is identical to the BRS variation.
Note also that $\dw v=0$ unlike eq.\ \nunn.

We can also find $\db$ algebra from eq.\ \brs\tfc\foo\fzt
\eqn\edbal{\eqalign{
&\db\CA =  - d_\CA v, \cr
&\db v =  -v^2, \cr
&\db\CF = -[v,\CF], \cr
&\db\Psi =  - [v,\Psi] , \cr
&\db \Phi =  -[v,\Phi],\cr}
}
which is identical to the additional BRS algebra of ref.\H.
We shall see that this additional BRS structure to $\dt$ and
$\dw$ is crucial for non-Abelian anomalies in TYM.

\appendix{B}{Solutions of Descent Equation}

Our conventions and the method to find the explicit solutions of
the extended descent equation \eedec\ are essentially same to those of
ref. \ZuminoB. However, for the sake of self-consistency, we will sketch
the procedure and present some explicit solutions.

We denote
\eqn\jack{
P(\hCF_1,\hCF_2,\ldots,\hCF_n),
}
to a symmetric invariant polynomial of degree $n$ in the Lie algebra
valued variables $\hCF_1,\hCF_2,\ldots,\hCF_n$. We shall write \jack\
as
\eqn\jake{
P(\hCF_1,\hCF_2,\hCF^{n-2})
}
for $\hCF_3=\hCF_4=\cdots=\hCF_n=\hCF$.

Note that if we introduce the one-parameter family of the total connection
one-forms
\eqn\susy{\eqalign{
\hCA_t &= t(\CA + C) + v\cr
&=t\,\hCA + v\cr
}}
and associated total field strengths
\eqn\susan{\eqalign{
\hCF_t &= (d +\dt +\db)\hCA_t + \hCA_t^2\cr
&\equiv (\hd +\db)\hCA_t + \hCA_t^2\cr
& = t\, \hd\hCA + t^2 \hCA^2 + (1-t)\,\hd v,
}}
we can obtain
\eqn\sarah{\eqalign{
P(\hCF_1^n) - P(\hCF_0^n)
&=n(\hd +\db)\int^1_0 dt\, P\left(\hCA,\,\hCF_t^{n-1}\right)\cr
&=n(\hd +\db)\int^1_0 dt\, P\left(\hCA,\,(t\,\hd\hCA + t^2\hCA^2 +(1-t)\,\hd
v)^{n-1}\right),\cr
}}
after
following the same procedure as eq.\ (2.15) - (2.17) in ref. \ZuminoB.
Now we can expand the right-hand side in powers of $\hd v$
\eqn\sanna{
n\int^1_0 dt\, P\left(\hCA,\,(t\,\hd\hCA + t^2\hCA^2 +(1-t)\,\hd
v)^{n-1}\right) = \CW_{2n-1}{}^0 +\CW_{2n-2}{}^1 +\cdots
\CW_{n}{}^{n-1},
}
where the superscript denotes the $\db$ ghost number and the subscript
denotes the space-time form degree plus the $\dt$-ghost number.

Then
\eqn\samanda{\eqalign{
\CW_{2n-1-k}{}^k &=\Fr{n(n-1)(n-2)\cdots(n-k)}{k!}\cr
&\times\int^1_0 dt\, (1-t)^k P((\hd v)^k,\,\hCA,\,(t\,\hd\hCA
+t^2\hCA^2)^{n-2}),\cr
}}
for $0\leq k\leq n-1$.
For $k >n-1$, we introduce a different family of the total connections
\eqn\june{
\hCA_t = t\, v,
}
and associated total field strengths
\eqn\jenny{
\hCF_t =(\hd +\db)\hCA_t +\hCA_t^2 = t\,\hd v + (t^2 -t)v^2.
}
After following the same procedure as eq.\ (2.27) - (2.30) of ref.
\ZuminoB, we can obtain
\eqn\jane{
\CW_{n-1-l}{}^{n+l} = (-1)^{l}\Fr{n!(n-1)!}{(n-1-l)!(n+l)!}
P((\hd v)^{n-1-l},v,(v^2)^{l}),
}
for $0\leq l\leq n-1$.

The next step is to expand $\CW_{2n-1-k}{}^k$, given by \samanda\ and
\jane, in powers of $\dt$ ghost number such that
\eqn\joan{
\CW_{2n-1-k}{}^{k} =\sum_{k^\prime =0}^{2n-1-k}
\CW_{2n-1-k-k^\prime}{}^{k,k^\prime}.
}
To be explicit, for $n=2$
\eqn\etwo{\eqalign{
%&\CW^0_3 = c_2 \tr\left(\hCA\,\hd\hCA+\f{2}{3}\hCA^3\right), \cr
&\CW_2{}^1 = c_2 \tr\left(\hd v\, \hCA \right), \cr
&\CW_1{}^2 = c_2 \tr\left(\hd v\,v \right), \cr
&\CW_0{}^3 = -\f{1}{3}c_2 \tr\left( v^3 \right),\cr }
}
we get
\eqn\eetwo{\eqalign{
%&\CW^{0,0}_3
%= c_2 \tr\left(\CA\,\CF - \f{1}{3}\CA^3\right),\cr
%&\CW^{0,1}_2
%= c_2 \tr\left(\CA\,(\Psi - \f{1}{3}\{\CA,C\})
%+ C(\CF -\f{1}{3} \CA^2\right),\cr
%&\CW^{0,2}_1
%= c_2 \tr\left(\CA\,(\Phi -\f{1}{3}C^2)
%+ C\,(\Psi  -\f{1}{3}\{\CA,C\}\right),\cr
%&\CW^{0,3}_0
%= c_2\tr\left(C\,\Phi -\f{1}{3}C^3\right),\cr
&\CW_2{}^{1,0}
= c_2\tr \left(dv\,\CA\right),\cr
&\CW_1{}^{1,1}
= c_2\tr \left(dv\, C + \dt v \,\CA\right),\cr
&\CW_0{}^{1,2}
= c_2\tr \left(\dt v\, C\right),\cr
&\CW_1{}^{2,0}
=  c_2\tr \left(dv\, v\right),\cr
&\CW_0{}^{2,1}
=  c_2\tr \left(\dt v\, v\right),\cr
&\CW_0{}^{3,0}_0
= -\f{1}{3}c_2 \tr  v^3.\cr}
}
For $n=3$
\eqn\ethr{\eqalign{
%&\CW^0_5
%= c_3\tr\left( \hCA\, (\hd \hCA)^2 +\f{3}{5}\hCA^5
%+\f{3}{2}\hCA^3\,\hd\hCA\right),\cr
&\CW_4{}^1
= \half c_3\tr\left(\hd v\,(\hCA\,\hd\hCA
+\hd\hCA\,\hCA +\hCA^3)\right),\cr
&\CW_3{}^2
= c_3\tr\left( (\hd v)^2\, \hCA \right),\cr
&\CW_2{}^3
= c_3\tr\left( (\hd v)^2\, v \right),\cr
&\CW_1{}^4
= -\half c_3\tr\left(\hd v\,v^3 \right),\cr
&\CW_0^5
=\f{1}{10}c_3\tr\left(v^5\right),\cr}
}
we find
\eqn\eethr{\eqalign{
%&\CW^{0,0}_5 =
%c_3\tr\left(\CA\,\CF^2-\half\CA^3\,\CF+\f{1}{10}\CA^5 \right),\cr
%&\CW^{0,1}_4 =
%c_3\tr\left(C\,\CF^2+2\CA\,\CF\,\Psi-\f{3}{2}\CA^2
%C\,\CF-\half\CA^3\,\Psi +\half\CA^4\,C \right),\cr
%&\CW^{0,2}_3 =
%c_3\tr\left(2\CA\,\CF\,\Phi+2C\,\CF\,\Psi-\f{3}{2}\CA\,C^2\,\CF
%+\CA\,\Psi^2-\half\CA^3\,\Phi-\f{3}{2}\CA^2\,C\,\Psi+\CA^3\,C^2\right),\cr
%&\CW^{0,3}_2 =
%c_3\tr\left(2C\,\CF\,\Phi-\half C^3\,\CF+2\CA\,\Phi\,\Psi+C\,\Psi^2
%-\f{3}{2}\CA^2\,C\,\Phi-\f{3}{2}\CA\,C^2\,\Psi+\CA^2\,C^3 \right),\cr
%&\CW^{0,4}_1 =
%c_3\tr\left(\CA\,\Phi^2+2C\,\Phi\,\Psi-\half C^3\,\Psi
%-\f{3}{2}\CA\,C^2\,\Phi+\half\CA\,C^4 \right),\cr
%&\CW^{0,5}_0
%=c_3\tr\left(C\,\Phi^2-\half C^3\,\Phi+\f{1}{10}C^5\right),\cr
&\CW_4{}^{1,0}
=\f{1}{2} c_3\tr\left(dv \left(\CA\,\CF +\CF\,\CA -\CA^3\right)\right),\cr
&\CW_3{}^{1,1}
=\f{1}{2}c_3\tr\left(dv\,
\left(\CA\,\Psi+\Psi\,\CA+C\,\CF+\CF\,C-\CA^2\,C-C\,\CA^2-\CA\,C\,\CA\right)
\right.\cr&\quad\qquad\qquad\left.
+ \dt v\,\left(\CA\,\CF +\CF\,\CA -\CA^3\right) \right),\cr
&\CW_2{}^{1,2}
=\f{1}{2}c_3\tr\left(dv
\left(C\,\Psi+\Psi\,C+\CA\,\Phi+\CA\,\Phi-\CA\,C^2-C^2\,\CA-C\,\CA\,C\right)
\right.\cr&\quad\qquad\qquad\left.
+\dt v\,
\left(\CA\,\Psi+\Psi\,\CA+C\,\CF+\CF\,C-\CA^2\,C-C\,\CA^2-\CA\,C\,\CA\right)
\right),\cr
&\CW_1{}^{1,3}
=\f{1}{2}c_3\tr\left(dv\,\left(C\,\Phi+\Phi\,C -C^3\right)
\right.\cr&\quad\qquad\qquad\left.
+\dt v\,
\left(C\,\Psi+\Psi\,C+\CA\,\Phi+\CA\,\Phi-\CA\,C^2-C^2\,\CA-C\,\CA\,C\right)
\right),\cr
&\CW_0{}^{1,4}
=\f{1}{2}c_3\tr\left(\dt v\,\left(C\,\Phi+\Phi\,C -C^3\right) \right),\cr
&\CW_3{}^{2,0}
=c_3\tr\left(\left(dv\right)^2\,\CA \right),\cr
&\CW_2{}^{2,1}
=c_3\tr\left(\left(dv\,\dt v+\dt v\,dv\right)\CA
+\left(dv\right)^2\,C\right),\cr
&\CW_1{}^{2,2}
=c_3\tr\left(\left(dv\,\dt v+\dt v\,dv\right)\,C
+\left(\dt v\right)^2\,C\right),\cr
&\CW_0{}^{2,3}
=c_3\tr\left(\left(\dt v\right)^2\, C\right),\cr
&\CW_2{}^{3,0}
=c_3\tr\left(\left(dv\right)^2\,v\right),\cr
&\CW_1{}^{3,1}
=c_3\tr\left(\left(dv\,\dt v+\dt v\,dv\right)\, v\right),\cr
&\CW_0{}^{3,2}
=c_3\tr\left(\left(\dt v\right)^2\,v\right),\cr
&\CW_1{}^{4,0}
=-\half c_3\tr\left(dv\,v^3\right),\cr
&\CW_0{}^{4,1}
=-\half c_3\tr\left(\dt v\, v^3\right),\cr
&\CW_0{}^{5,0}
= \f{1}{10}c_3\tr\left(v\right)^5.\cr}
}
\vfill
\eject
\listrefs
\end
\bye